\documentclass[preprint,groupedaddress,showkeys,showpacs]{revtex4}
\usepackage{latexsym, amssymb, amsmath}
\usepackage{graphicx}

\newcommand{\PDfrac}[2]{ \frac{\partial\,#1}{\partial\,#2} }
\newcommand{\Dfrac}[3][{}]{ \frac{d^{\,#1}#2}{d\,#3^{#1}} }

\begin{document}

\title{Mathematical model for blood flow autoregulation by 
endothelium-derived relaxing factor}

\author{I.L. Chernyavsky}
\author{N.A. Kudryashov}
\email{kudryashov@mephi.ru}
\affiliation{Department of Applied Mathematics,\\
Moscow  Engineering and Physics Institute (State University)\\
31 Kashirskoe Shosse,  115409, Moscow, Russian Federation}

\begin{abstract}
The fluid shear stress is an important regulator of the
cardiovascular system via the endothelium-derived relaxing factor
(EDRF) that is Nitric Oxide. This mechanism involves biochemical
reactions in an arterial wall. The autoregulation process is managed
by the vascular tonus and gives the negative feedback for the shear
stress changing. A new mathematical model for the autoregulation of
a blood flow through arteria under the constant transmural pressure
is presented. Endothelium-derived relaxing factor Nitric Oxide, the
multi-layer structure of an arterial wall, and kinetic-diffusion
processes are taken into consideration. The limit case of the
thin-wall artery is analytically studied. The stability condition
for a stationary point of the linearized system is given. The exact
stationary solutions of the origin system are found. The numerical
simulation for the autoregulation system is presented. It is shown
the arteria adaptation to an initial radial perturbation and the
transition of the system to new equilibrium state in response on the
blood flow changing.
\end{abstract}

\pacs{87.10.+e, 87.15.Rn, 87.15.Vv, 87.16.Ac, 87.19.Uv}
\keywords{arteria, blood flow, autoregulation, EDRF, Nitric Oxide,
kinetic-diffusion process, viscoelasticity}

\maketitle

\section{Introduction}
The modeling of a cardiovascular system is the problem of great importance due
to it should be aid for understanding and prediction of various diseases like
atherosclerosis, arteriosclerosis, hypertension, etc. The most often approaches
to the problem is a direct applying of classical hydrodynamics models of fluid
flow through elastic shells or
tubes~\cite{Canic2003, Pontrelli2003, Quarteroni2003}.
But in some cases, for example, for muscular resistance arteries, it is
necessary to take into account the difference between an ordinary passive and
a ''biological'' active tube.

The one way to reach the purpose is a wide discussed effect of the
flow-induced vasodilation by Nitric Oxide (NO)
radical~\cite{Snow_etal2001, Buchanan1993, Rachev2000, Smith2003}.
For a long time the endothelium cells, covering the arterial bed
surface, was supposed to provide only the friction reduction for the
blood flow through arteria. But since the EDRF was discovered via
the comparison of two arterial rings with and without endothelium by
the ability to acetylcholine-dependent smooth muscle relaxation and
due to later investigations it was shown the endothelium plays the
main role for the local blood flow regulation. The so-called
Endothelium Derived Relaxing Factor (EDRF) was discovered in 1980 by
Robert~F.~Furchgott~\cite{Furchgott1980,Furchgott1999}. Nitric oxide
was proposed as a signal molecule to set a connection from
endothelium to the smooth muscles. The EDRF-NO mechanism enabled to
explain the principle of action for the first-aid medicine
Nitroglycerine being used before without understanding.

The mechanical nature of an arterial wall tonus regulation is highlighted in the
works~\cite{Payne2005, Rachev2000, Joannides_etal1995, Rubanyi1986}.
It is perceived that the increasing of a shear stress between the blood flow
and inner arterial surface causes the relaxation in a smooth muscle
layer of an arterial wall. This necessary induces the increasing of the
arterial radius and the decreasing of the shear stress itself. Therefore the
process in a whole gives the system with a negative feedback.

There are three main layers in an arterial wall.
The first, internal layer is intima~(i), the second layer is media~(m) and
the last one is adventitia~(a). The inner boundary of the intima layer, i.e.
the internal arterial surface, is covered with endothelium cells. The media
layer is full of smooth muscle cells. The thicknesses of the layers depend on
the type of artery or arteriole. We mainly consider the muscle resistance
arteries which have well developed muscle layer (media) and non-vanishing
intima layer. The typical ratio of intima thickness to the media one is about
$10^{-1}$.

The scenario of the flow-induced relaxation is as follows. The increasing shear
stress $\sigma_{shear}$ on the surface of endothelium cells opens calcium
channels which launch the production of NO from L-arginine with NO-synthase
(NOS) catalysis and then NO diffuses with descending through the intima
layer towards the smooth muscle cells in the media layer. As a lipophilic
molecule NO easily penetrates through a cell membrane of a muscle cell and
initiate synthesis of the cyclic guanosine monophosphate (cGMP). Ultimately,
cGMP stimulates the outflow of intracellular $Ca^{2+}$ that leads to relaxation
of the smooth muscle cell. The flow-induced contraction is realized vise versa.

The aim of this paper is to develop and to study the mathematical
model for description of blood flow autoregulation that accumulates
a viscoelastic nature of an arterial wall and the two-layer
diffusion and kinetic processes for concentrations of the key
agents: Nitric Oxide (NO) and Calcium ions ($Ca^{2+}$).

The outline of the article is as follows. In the
section~\ref{sec:Main_assumptions} we introduce the assumptions of
the model. In the section~\ref{sec:Math_model_derivation} we derive
the closed system for description of the autoregulation process. In
the section~\ref{sec:Steady_state} the steady-state concentrations
of NO and $Ca^{2+}$ are obtained. In the
section~\ref{sec:Thin-wall_artery} we study the limit case of a
thin-wall artery. In this case the stability condition of an
equilibrium state of the system is given. In the
section~\ref{sec:Passive_tube} we consider the case of passive
dilation of an arteria with the fully relaxed muscles. The exact
kink-shaped solution is found. In the 
section~\ref{sec:Numerical_simulation} the numerical simulation
of the autoregulation process near the stationary state is presented.
In the section~\ref{sec:Conclusion} we summarize and discuss the 
obtained results. The appendix~\ref{sec:Appendix-notation} includes 
the essential notations. The appendix~\ref{sec:Appendix-SimpleEq-method} 
gives the approach for finding exact solution of the passive vessel model.

\section{Main assumptions of the model}\label{sec:Main_assumptions}
We consider the arteria to be axial-symmetric, viscoelastic and incompressible.
The blood is also assumed to be incompressible and Newtonian. The flow is
quasi-stationary, the transmural pressure is constant and the velocity profile
is the power generalization of the Poiseuille's law. We suppose the dependence
of muscular force on calcium concentration to be linear and the dependence of
$Ca^{2+}$ concentration decreasing ratio on Nitric Oxide concentration is also
linear. The concentration of NO in the endothelium is assumed to be
proportional to the shear stress on an arterial wall~\cite{Rachev2000}.

\section{The statement of the problem}\label{sec:Math_model_derivation}
Let us consider an arterial segment of length $l$ in the cylindrical
coordinate system $( r, \theta, x \equiv z )$. The intima, media, and
adventitia layers have coordinates $R_i,\,R_m,\,R_a$ respectively.

\subsection{The shear stress dependence on the blood flow}

Consider the power generalization velocity profile of the Poiseuille's
flow~\cite{Quarteroni2003}:
\begin{equation}
  V_{x}(r,x,t) = \frac{\gamma+2}{\gamma}
    \left[\, 1 - \left(\frac{r}{R(t)}\right)^{\gamma}\, \right]\,\bar{u}(x,t)
\end{equation}
Here $V_{x}$ is the axial velocity, $\bar{u}$ is the cross-sectional averaged
axial velocity, $R$ is the arterial radius and $\gamma$ is the profile
sharpness.

In case of Newtonian fluid with dynamical viscosity $\mu$ the shear stress on
the wall of elastic tube is
\begin{equation}
  \sigma_{shear} = -\mu \left.\PDfrac{V_{x}}{r}\right|_{r=R}
      = (\gamma + 2)\mu \frac{\bar{u}}{R} = (\gamma + 2)\mu\,\frac{Q}{\pi R^3}
  \label{eq:shear_stress}
\end{equation}
where $Q = A\bar{u}$ is the blood discharge through the cross-section with
the area $A$.

Summarize all the assumptions for the laminar stationary flow,
the cross-section averaged Navier-Stokes equation takes the form of the
generalized Hagen-Poiseuille equation:
\begin{equation}
  \frac{\Delta P}{l} = 2(\gamma + 2)\mu \frac{Q}{\pi R^4}
  \label{eq:generalized_Poiseuille}
\end{equation}
where $\Delta P$ is the pressure difference on an arterial segment with the
length $l$.

It shows the linear dependence of the pressure gradient from the discharge and
inversely proportionality to the fourth power of the arterial radius.

In case of axial-symmetric radial perturbations $R(t) = R_{0} + \eta(t)$
we have from (\ref{eq:shear_stress}):
\begin{equation}
  \sigma_{shear} = \frac{ (\gamma + 2)\mu }{ \pi R_{0}^3 }\,
    \frac{ Q }{ \left(1 + \dfrac{\eta}{R_0}\right)^3 }
  \label{eq:shear_stress_small_pert}
\end{equation}

There is a hypothesis of maintaining the shear stress principle
$\sigma_{shear} = const$ \cite{Togawa1980,Rachev2000}. One can conclude the
increasing of the flow necessitate the increasing of the steady-state arterial
radius to compensate the changing of the shear stress. The estimated relation
between the new steady-state discharge and new stationary radius is as follows:
\begin{equation}
  \eta = \left( \sqrt[3]{ \frac{Q}{Q_{0}} } - 1 \right) R_{0}
\end{equation}

It is remarkable the difference of reaction for increasing and decreasing the
blood flow near the previous stationary value. The changing of the radius in
response to higher flow is smaller than for the same lower flow. It is
explained by the inversely cubic dependence of shear stress from the radius.

One can see there is the linear dependence between radial perturbation and
mean blood flow in case of small radial perturbations ($|\eta| << R_{0}$).

\subsection{The synthesis and diffusion of Nitric Oxide}

According to the EDRF-mechanism mediated by the fluid flow, the
concentration of Nitric Oxide produced by an endothelium cell is
managed by the shear stress value. We consider the NO transport to
the smooth muscle tissue as a diffusion process (diffusion
coefficient is $D_1$) with a descending (reaction coefficient is
$\delta_1$). Then it continues to diffuse through the media layer
but with another diffusion coefficient $D_2$ and reaction
coefficient $\delta_2$. 

The production of Nitric Oxide in an endothelium cell has the shear 
stress $\sigma_{shear}$ as one of essential regulators, therefore this 
process can be described with a kinetic equation:
\begin{equation}
  \Dfrac{n_{e}}{t} = - k_{e}\,n_{e} + \psi\,\sigma_{shear}(t)
  \label{eq:NO_production}
\end{equation}
where $n_{e}$ is the NO concentration in an endothelium cell, 
$k_{e}$ is the rate of mass transfer of NO from the cell, and
$\psi$ is the production rate constant.

Under the assumption of quasi-stationary NO production, i.e.
that characteristic time of NO mass transfer from an endothelium cell
towards intima layer ($\tau_{NO-mass-transfer} \sim \Delta r^2/D \simeq 1/528\, {sec}$,
where $D=3300\,{\mu}m^2/sec,\; \Delta r \simeq 2.5\,{\mu}m$ 
\cite{Regirer2005}) is smaller than the typical 
time of $\sigma_{shear}$ changing ($\tau_{shear} \sim 
\tau_{radius-oscillations} \sim 1/2 \,{sec}$), from equation 
(\ref{eq:NO_production}) we have the following relation between 
the $n_{e}$ and $\sigma_{shear}$:
\begin{equation}
  n_{e}(t) = \frac{\psi}{k_{e}}\,\sigma_{shear}(t) 
  \label{eq:inner_bound_condition_for_NO}
\end{equation}

The relation (\ref{eq:inner_bound_condition_for_NO}) is used
as inner boundary condition for Nitric Oxide diffusion through
an arterial wall ($n|_{r=R_{intima}} = n_{e}$).
Ultimately, at the inner boundary of intima layer we assume the
concentration of NO to be proportional to the shear stress
(proportionality coefficient is $k_{3}$). Between the intima and
media layers we use the continuity of concentrations and fluxes. On
the external layer we take the impenetrability condition into
account. Thus the system of equations and the boundary conditions
for the Nitric Oxide concentration are as follows:
\begin{equation}
  \begin{gathered}
    \PDfrac{ n_{j} }{ t } =  D_{j} \frac{1}{r}\PDfrac{}{r}
          \left(r\, \PDfrac{n_{j}}{r} \right) - \delta_{j}\, n_{j},  \hfill  \\
          R_{i} < r < R_{m}\quad \mbox{for}\; j=1\quad \mbox{(intima)} \hfill\\
          R_{m} < r < R_{a}\quad \mbox{for}\; j=2\quad \mbox{(media)}  \hfill\\
    n_{1}|_{r=R_{i}} = k_{3}\,\sigma_{shear} \hfill \\
    n_{1}|_{r=R_{m}} = n_{2}|_{r=R_{m}},\quad
         \left. D_{1}\PDfrac{n_{1}}{r} \right|_{r=R_{m}} =
         \left. D_{2}\PDfrac{n_{2}}{r} \right|_{r=R_{m}}  \hfill \\
    \left. \PDfrac{n_{2}}{r} \right|_{ r=R_{a} } = 0  \hfill
  \label{eq:NO_diffusion}
  \end{gathered}
\end{equation}

The system of equations (\ref{eq:NO_diffusion}) together with
initial conditions describes the two-layer diffusion-kinetic process
for the Nitric Oxide in an arterial wall.

\subsection{The equation for the kinetics of the Calcium ions in a smooth
muscle cell}

To derive the balance-equation for concentration of $Ca^{2+}$ in a smooth
muscle cell it is necessary to describe the ways of $Ca^{2+}$ in- and
out-fluxes. There are two source of the calcium ions: the extracellular space
and the intracellular containers -- sarcoplasmic reticulum. The concentration of
$Ca^{2+}$ in these sources is about $10^4$ greater than in the intracellular space.
The balance of calcium ions in the muscle cell at the point $r$ may be
described, similarly to \cite{Rachev2000}, as
\begin{equation}
  \PDfrac{C(r,t)}{t} = -\alpha(C - C_{0}) + \beta(C_{ext} - C) - k_{1}n_{2}(r,t)
  \label{eq:full_Ca-balance_eq}
\end{equation}
where the first term is responsible for the natural active outflow transport of
$Ca^{2+}$ compared to the minimal observed concentration $C_{0}$, the second
term is described a passive diffusion provided by the difference between the
intracellular calcium concentration $C$ and extracellular ones
$C_{ext}$, and the last term is presented the NO-mediated active outflow.

Taking into account the relation $C_{ext} >> C$ we can treat it as a
constant source: $\varphi_{0} = \alpha C_{0} + \beta C_{ext} =
const$. In this case the equation (\ref{eq:full_Ca-balance_eq}) can
be transformed to the form:
\begin{equation}
  \PDfrac{C(r,t)}{t} = -\alpha C - k_{1} n_{2}(r,t) + \varphi_{0}
  \label{eq:Ca-balance_eq}
\end{equation}

The equation (\ref{eq:Ca-balance_eq}) is used to describe the
Calcium-balance in the smooth muscle layer.

\subsection{The equation for an arterial wall movement}

In order to obtain the close system of a blood flow autoregulation we need to
have a link between the radial perturbation and the external forces such as
pressure and muscular force \cite{ChernKudr2006}. The constitutive
equation~\cite{FungBook1993} can be found from the movement equation for
an arterial wall segment.

Let us consider the incompressible viscoelastic wall element with mass
$\Delta m$, density $\rho_{w}$, width $h$, radius $R$, and length $\Delta x$.
According to the movement law
\begin{equation}
  \begin{gathered}
    \Delta m \Dfrac[2]{R}{t} = f_{radial} + f_{pressure}, \\
    f_{radial} = -\sigma_{\theta\theta}\, 2 \pi \Delta x h,\quad
    f_{pressure} = (\bar{P} - P_{ext})\, 2 \pi \Delta x h
  \end{gathered}
  \label{eq:general_wall_mov_eq}
\end{equation}
where $\Delta m = \rho_{w} 2 \pi R \Delta x h$, $f_{radial}$ is proportional
to the circumference component of a stress tensor $\sigma_{\theta\theta}$
and $f_{pressure}$ is the resulting transmural pressure (the difference
between the internal and external pressure).

The stress tensor component $\sigma_{\theta\theta}$ consists of three parts:
a passive elastic force (weakly nonlinear with quadratic addition),
a viscous force and an active force due to the muscle tonus
\begin{equation}
  \sigma_{\theta\theta} = \frac{E(<\!C\!>)}{1-\xi^2}\frac{R-R_0}{R_0}
    + E_{1} \left( \frac{R-R_0}{R_0} \right)^2
    + \lambda \Dfrac{R}{t} + k_{2}\,F(C)
  \label{eq:stress_tensor}
\end{equation}
here $E(<\!C\!>)$ is the Yung's modulus dependent on averaged concentration of
$Ca^{2+}$ in a muscle cell layer, $\xi$ is the Poisson's ratio,
$E_{1}$ is the small nonlinear elastic coefficient for a square
addition, $\lambda$ is the viscous characteristic of the wall, $F(C)$ is
the active force component determined also by the integral calcium
concentration level above the threshold one $C_{th}$, $k_{2}$ is
the coefficient of proportionality for the muscular tonus response on the
$Ca^{2+}$ level.

Substitute (\ref{eq:stress_tensor}) in (\ref{eq:general_wall_mov_eq}) and
take into consideration the linear dependence of muscle force on calcium and
the incompressibility condition $h_{0}R_{0} = h R$. Then the constitutive
equation for the radial perturbations (${R = R_{0} + \eta}$,\,
${|\eta| << R_{0}}$) has a form
\begin{equation}
  \begin{gathered}
  \rho_{w}h_{0}\Dfrac[2]{\eta}{t} + \frac{\lambda h_{0}}{R_{0}}\,\Dfrac{\eta}{t}
    + \frac{\varkappa(C)h_{0}}{R_{0}}\,\eta
    + \frac{E_{1}h_{0}}{R_{0}^{3}}\,\eta^2 = \\
      \hfill  = (\bar{P} - P_{ext}) - \frac{h_{0}}{R_{0}}\, k_{2}\, F(C)
  \end{gathered}
\end{equation}
where
\begin{equation}
  \begin{gathered}
    \varkappa(C) = \varkappa_{0}(1 + \varepsilon F(C)),\quad
    \varkappa_{0} = \frac{E_{0}}{R_{0}(1-\xi^2)},\quad \varepsilon << 1  \\
    F(C) = \int_{R_m}^{R_a} [\,C - C_{th}\,]\,\theta(C - C_{th})\, r\,dr\,,
      \hfill \\
      \theta\, \mbox{-- the Heaviside's step function} \hfill
  \end{gathered}
\end{equation}

Renormalize the constants $\lambda, \varkappa, k_2$ with the value $h_0/R_0$
and denote the constant, under the assumptions, transmural pressure
$P_{0} = \bar{P} - P_{ext} = const$ and
${\varkappa_{1} = \frac{E_{1}h_{0}}{R_{0}^{3}}}$.

Ultimately, we obtain a new integro-differential equation describing
the wall movement in the presence of smooth muscle tonus
\begin{equation}
  \begin{gathered}
     \rho_{w}h_{0} \Dfrac[2]{\eta}{t} + \lambda\,\Dfrac{\eta}{t}
       + \varkappa(C)\,\eta + \varkappa_1\,\eta^{2} = P_{0} 
         -\,k_{2} \int_{R_m}^{R_a} [\,C - C_{th}\,]\,\theta(C - C_{th})\, r\,dr
  \end{gathered}
  \label{eq:consitutive_eq}
\end{equation}

One can see in the case of absence of muscle force (full relaxation) it is the
equation of a nonlinear damping oscillator with an external force. The presence
of calcium-dependent force term is provided the feedback and makes the arteria
different from a passive viscoelastic tube.

\subsection{The problem statement for the blood flow autoregulation
in dimensionless variables}

Summarize the equations obtained above we have the complete system to
describe the process of blood flow autoregulation due to EDRF-NO mechanism:
\begin{equation}
  \PDfrac{C(r,t)}{t} = - \alpha\, C - k_{1}\,n_{2}(r,t) + \varphi_{0}, \quad
          R_{m} < r < R_{a}
  \label{eq:Ca_balance}
\end{equation}
\begin{equation}
  \PDfrac{ n_{1} }{ t } =  D_{1} \frac{1}{r}\PDfrac{}{r}\left(r\, \PDfrac{n_{1}}{r} \right)
          - \delta_{1}\, n_{1},\quad    R_{i} < r < R_{m}
  \label{eq:NO_diffusion_in_intima}
\end{equation}
\begin{equation}
  \PDfrac{ n_{2} }{ t } =  D_{2} \frac{1}{r}\PDfrac{}{r}\left( r\, \PDfrac{n_{2}}{r} \right)
          - \delta_{2}\, n_{2},\quad    R_{m} < r < R_{a}
  \label{eq:NO_diffusion_in_media}
\end{equation}
\begin{equation}
  \begin{gathered}
     \rho_{w}\, h_{0}\, \Dfrac[2]{\eta}{t} + \lambda \, \Dfrac{\eta}{t} + \varkappa(C)\, \eta
          + \varkappa_1\, \eta^{2} = P_{0} 
          -\, k_{2} \int_{R_m}^{R_a} [\,C - C_{th}\,]\,\theta(C - C_{th})\, r\,dr
  \end{gathered}
  \label{eq:wall_movement_eq}
\end{equation}
with the boundary conditions:
\begin{equation}
  \begin{gathered}
    n_{1}|_{r=R_{i}} = k_{3}\,\sigma_{shear}
    = \frac{ k_{3}\,(\gamma + 2) \mu\, Q }
           { \pi\,R_0^{3} \left( 1 + \frac{\eta}{R_0} \right)^3 }  \hfill \\
    n_{1}|_{r=R_{m}} = n_{2}|_{r=R_{m}},\quad
         \left. D_{1}\, \PDfrac{n_{1}}{r} \right|_{r=R_{m}} =
         \left. D_{2}\, \PDfrac{n_{2}}{r} \right|_{r=R_{m}}  \hfill \\
    \left. \PDfrac{n_{2}}{r} \right|_{ r=R_{a} } = 0  \hfill
  \end{gathered}
  \label{eq:boundary_conditions}
\end{equation}

As the initial values it is taken the perturbed steady-state solutions.

Here equation (\ref{eq:Ca_balance}) describes the $Ca^{2+}$-balance in a
smooth-muscle cell, equations
(\ref{eq:NO_diffusion_in_intima}),~(\ref{eq:NO_diffusion_in_media})
characterize the diffusion of Nitric Oxide in intima and media respectively,
and the equation (\ref{eq:wall_movement_eq}) gives the relation establishing
the arterial wall movement under the influence of the average calcium ions
concentration.

In order to pass on to the non-dimensional system of equation setting up the new
dimensionless variables:
\begin{equation}
  \begin{gathered}
    C = C_{th}\, C',\quad  n_{1} = n_{1}^{0}\, n_{1}',\quad  n_{2} = n_{2}^{0}\, n_{2}', \\
    \eta = \eta_{0}\, \eta',\quad  t = t_{0}\, {t}',\quad  r = r_{0}\, {r}'
  \end{gathered}
  \label{eq:non-dim_variables}
\end{equation}
were for convenience choosing
\begin{equation}
  \begin{gathered}
    n^{0} \equiv n_{1}^{0} = \frac{D_{2}}{D_{1}}n_{2}^{0}
      = k_{3}\,\sigma_{shear}^{0}
      = \frac{ k_{3}\,(\gamma + 2) \mu\, Q }{ \pi\,{R_0}^{3} } \\
    r_0 = \eta_{0} = R_0,\quad  t_{0} = \frac{1}{\alpha},\quad  R_0 = R_{i}\\
  \end{gathered}
\end{equation}

After substitution (\ref{eq:non-dim_variables}) the system
(\ref{eq:Ca_balance})~--~(\ref{eq:wall_movement_eq}) turns into a dimensionless
form (primes over the variables are omitted):
\begin{equation}
  \PDfrac{C}{t} = - C - k'_{1}\,n_{2} + \varphi'_{0}, \quad R'_{m} < r < R'_{a}
  \label{eq:non-dim_Ca_balance}
\end{equation}
\begin{equation}
  \PDfrac{n_{1}}{t} =  D'_{1}\frac{1}{r}
    \PDfrac{}{r}\left( r\, \PDfrac{n_{1}}{r} \right)
    - \delta'_{1}\, n_{1}, \quad    1 < r < R'_{m}
  \label{eq:non-dim_NO_diffusion_in_intima}
\end{equation}
\begin{equation}
    \PDfrac{n_{2}}{t} =  D'_{2}\frac{1}{r}
    \PDfrac{}{r}\left( r\, \PDfrac{n_{2}}{r} \right)
    - \delta'_{2}\, n_{2}, \quad    R'_{m} < r < R'_{a}
  \label{eq:non-dim_NO_diffusion_in_media}
\end{equation}
\begin{equation}
  \begin{gathered}
    \Dfrac[2]{\eta}{t} + \lambda'\, \Dfrac{\eta}{t} + \varkappa'\,\eta
      + \varkappa_{1}'\, \eta^{2} = P'_{0} 
      -\, k'_{2} \int_{R'_m}^{R'_a} [\,C - 1\,]\; \theta(C - 1)\, r\,dr
  \end{gathered}
  \label{eq:non-dim_wall_movement_eq}
\end{equation}
where dimensionless constants are
\begin{equation}
  \begin{gathered}
    k'_{1} = \frac{ k_{1}\,n^{0} }{ \alpha\, C_{th} },\quad
    \varphi'_{0} = \frac{ \varphi_{0} }{ \alpha\, C_{th} }
                        \equiv  \frac{ \beta\, C_{ext} }{ \alpha\, C_{th} },  \hfill \\
    D'_{1,2} = \frac{ D_{1,2} }{ \alpha\,{R_0}^{2} },\quad
    \delta'_{1,2} = \frac{ \delta_{1,2} }{ \alpha },\quad
    \lambda' = \frac{ \lambda }{ \alpha\,\rho_{w}\, h_{0} },\quad  \hfill \\
    \varkappa' = \frac{ \varkappa }{ \alpha^{2}\,\rho_{w}\, h_{0} },\quad
    \varkappa'_{1} = \frac{ \varkappa_{1}\,R_{0} }{ \alpha^{2}\,\rho_{w}\, h_{0} },
      \hfill \\
    P'_{0} = \frac{ P_{0} }{ \alpha^{2}\,\rho_{w}\, h_{0}\,R_0 },\quad
    k'_{2} = \frac{ k_{2}\,C_{th}\,R_0 }{ \alpha^{2}\,\rho_{w}\, h_{0} }  \hfill
  \end{gathered}
\end{equation}

Then the boundary conditions take the form:
\begin{equation}
  \begin{gathered}
    n_{1}|_{r=1} = \frac{ 1 }{ ( 1 + \eta )^{3} }  \hfill \\
    n_{1}|_{r=R'_{m}} = n_{2}|_{r=R'_{m}},\quad
         \left. \PDfrac{n_{1}}{r} \right|_{r=R'_{m}} =
         \left. \PDfrac{n_{2}}{r} \right|_{r=R'_{m}}  \hfill \\
    \left. \PDfrac{n_{2}}{r} \right|_{r=R'_{a}} = 0  \hfill
  \end{gathered}
  \label{eq:non-dim_boundary_conditions}
\end{equation}
where $R_{0}=R_{i},\quad R'_{m}=R_{m}/R_0,\quad R'_{a}=R_{a}/R_0$\quad
and initial values are the perturbed solutions of the steady-state system.

From the non-dimensional system of equations one can make a remark that the
stationary blood flow discharge through the vessel's cross-section $Q$ has
implicit influence on the $Ca^{2+}$-concentration in the smooth muscle cell
via term $k'_{1}\, n_{2}$ in the equation (\ref{eq:non-dim_Ca_balance})
due to the coefficient $k'_{1} \sim n^{0} \sim Q$.

\section{The solution of the problem in a steady state}
\label{sec:Steady_state}
To consider the stationary case letting the following:
\begin{equation}
  C = \tilde{C}(r),\; n_{1} = \tilde{n}_{1}(r),\; n_{2} = \tilde{n}_{2}(r),\;
  R = R_{0} = const
\end{equation}

Under the assumptions the system of equations
(\ref{eq:non-dim_Ca_balance})~--~(\ref{eq:non-dim_wall_movement_eq})
takes the form:
\begin{equation}
  \tilde{C}(r) = - k'_{1}\,\tilde{n}_{2}(r) + \varphi'_{0},\quad
  R'_{m} \leq r \leq R'_{a}
  \label{eq:steady-state_Ca}
\end{equation}
\begin{equation}
  \Dfrac[2]{\tilde{n}_{1}}{r} + \frac{1}{r}\,\Dfrac{\tilde{n}_{1}}{r}
  - \frac{ \delta'_{1} }{ D'_{1} }\, \tilde{n}_{1} = 0,\quad  1 \leq r \leq R'_{m}
  \label{eq:steady-state_NO_in_intima}
\end{equation}
\begin{equation}
  \Dfrac[2]{\tilde{n}_{2}}{r} + \frac{1}{r}\,\Dfrac{\tilde{n}_{2}}{r}
  - \frac{ \delta'_{2} }{ D'_{2} }\, \tilde{n}_{2} = 0,\quad  R'_{m} \leq r \leq R'_{a}
  \label{eq:steady-state_NO_in_media}
\end{equation}
\begin{equation}
  P'_{0} = k'_{2}\int_{R'_m}^{R'_a}
               [\,\tilde{C}(r) - 1\,]\,\theta(\tilde{C} - 1)\, r\,dr
  \label{eq:steady-state_wall_movement_eq}
\end{equation}

with the boundary conditions:
\begin{equation}
  \begin{gathered}
    \tilde{n}_{1}|_{r=R'_{i}} = 1 \hfill \\
    \tilde{n}_{1}|_{r=R'_{m}} = \tilde{n}_{1}|_{r=R'_{m}},\quad
         \left. \Dfrac{\tilde{n}_{1}}{r} \right|_{r=R'_{m}} =
         \left. \Dfrac{\tilde{n}_{2}}{r} \right|_{r=R'_{m}} \hfill \\
    \left. \Dfrac{\tilde{n}_{2}}{r} \right|_{r=R'_{a}} = 0  \hfill
  \end{gathered}
  \label{eq:steady-state_bound_cond}
\end{equation}

The ODEs
(\ref{eq:steady-state_NO_in_intima}),~(\ref{eq:steady-state_NO_in_media}) for
NO concentration have general solution via modified Bessel functions
$I_0(z), K_0(z)$:
\begin{equation}
  \begin{gathered}
    \tilde{n}_{1}(r) = A_1\; I_{0}\!\left( \sqrt{ \frac{\delta'_1}{D'_1} }\, r \right)
                     + A_2\; K_{0}\!\left( \sqrt{ \frac{\delta'_1}{D'_1} }\, r \right) \\
    \tilde{n}_{2}(r) = B_1\; I_{0}\!\left( \sqrt{ \frac{\delta'_2}{D'_2} }\, r \right)
                     + B_2\; K_{0}\!\left( \sqrt{ \frac{\delta'_2}{D'_2} }\, r \right)
  \end{gathered}
  \label{eq:exact_steady-state_NO-distr}
\end{equation}
where $A_1,\, A_2, B_1,\, B_2$ are the arbitrary constants defining by
the boundary conditions (\ref{eq:steady-state_bound_cond}):
\begin{equation}
  \begin{gathered}
    A_1\, I_{0}(\xi_1) + A_2\, K_{0}(\xi_1) = 1  \hfill \\
    B_1\, I_{1}(\xi_2\,R'_{a}) - B_2\, K_{1}(\xi_2\,R'_{a}) = 0   \hfill \\
    A_1\, I_{0}(\xi_1\,R'_{m}) + A_2\, K_{0}(\xi_1\,R'_{m}) =     \hfill \\
      \hfill = B_1\, I_{0}(\xi_2\,R'_{m}) + B_2\, K_{0}(\xi_2\,R'_{m})          \\
    \xi_1\, (A_1\, I_{1}(\xi_1\,R'_{m}) - A_2\, K_{1}(\xi_1\,R'_{m}) ) = \hfill \\
      \hfill = \xi_2\, ( B_1\,I_{1}(\xi_2\,R'_{m}) - B_2\, K_{1}(\xi_2\,R'_{m}) )
  \label{eq:bound_cond_system}
  \end{gathered}
\end{equation}
where $\xi_1 \equiv \sqrt{ \frac{\delta'_1}{D'_1} },\quad
       \xi_2 \equiv \sqrt{ \frac{\delta'_2}{D'_2} }$

Using the typical experimental data for the muscular resistance artery
~\cite{Dorf2003, Li2004, Regirer2005}\\
$R_{i} = 1.0\, mm,\: h=0.5\, mm$; $R'_{m}=1.05,\: R'_{a}=1.3$
and assuming $\xi_{1}=6,\: \xi_{2}=2$ we can find the constants
$A_1,\, A_2,\, B_1,\, B_2$ from the boundary
conditions~(\ref{eq:bound_cond_system}).

Steady-state $Ca^{2+}$- concentration $\tilde{C}(r)$ is given by
(\ref{eq:steady-state_Ca}).

\begin{figure}[!ht]
  \centering              
    \includegraphics[width=9cm]{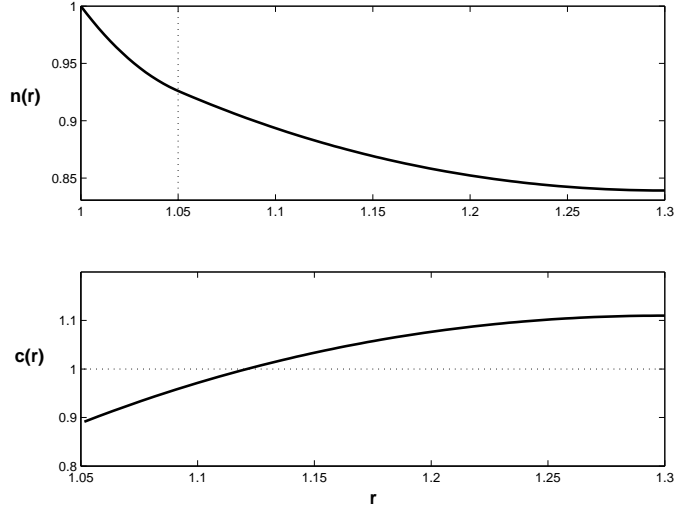}
    \caption{Stationary distribution of NO and intracellular Calcium ions.}
  \label{fig:stationary-case}
\end{figure}

The equilibrium distribution of concentrations is depicted on
the figure~\ref{fig:stationary-case}.

\section{The case of a thin-wall artery}\label{sec:Thin-wall_artery}
To understand the qualitative behavior of the system consider the limit case
of a thin wall artery. The similar model was studied by A.~Rachev,
S.A.~Regirer et al. in \cite{Rachev2000, Regirer2002}.
There are estimations to come to the limit case. The first relation is
$h_{i}/h_{m} << 1$ that enables to come to a one-layer wall model.
The second one is $\tau_{diffusion} << \tau_{kinetic}$,
were the typical time of the diffusion process is
$\tau_{diffusion} = \frac{h^2}{D}$ and the typical time of the kinetic process
is $\tau_{kinetic} = min\{ \frac{1}{\delta},\, \frac{1}{\alpha} \}$.
Here $h_{i}, h_{m}$ are the wall thickness of intima and media layers,
$h$ is the spatial scale of the wall thickness. Considering that the kinetic
processes for Nitric Oxide are faster than for Calcium ions we have as follows:
\begin{equation}
  h << \sqrt {\frac{D}{\delta}} \equiv h_{0}
  \label{eq:thickness_estimation}
\end{equation}
where $h_{0}$ is the characteristic wall thickness to compare with.
Taking into account the typical values of the parameters as
$D = 3300\, \mu{m}^2/sec$ and $\delta = 1\, sec^{-1}$ \cite{Regirer2005}
we obtain $h_{0} = 57\, \mu{m}$.

It is also should be note the default condition of quasi-stationary diffusion:
$\tau_{diffusion} << T_{osc}$, where $T_{osc}$ is the typical period of radial
oscillations. The typical value of $T_{osc}^{-1}$ is about $1 \div 2\,sec^{-1}$
then the $h_{0}$ value is close to $57\,\mu{m}$ or a bit less.

The large and medium resistance muscle arteries have the specific wall
thickness ${h \sim 100 \div 1000\, \mu{m}}$ whereas the small arteries and
arterioles have the much smaller thickness ${h \sim 10\, \mu{m}}$. Therefore the
limit case covers the case of flow in a small artery with ${h << 50\, \mu{m}}$.

Thus the intima and media layers are so thin to neglect the multi-layer nature
of the wall and eliminate the diffusion processes.

After the averaging of the calcium feedback $F(C)$ over the wall thickness the
system (\ref{eq:Ca_balance})~--~(\ref{eq:wall_movement_eq}) takes a simplified form:
\begin{equation}
  \begin{gathered}
    \Dfrac{x}{t} = - \alpha\,x - \dfrac{a}{\left(1 + \frac{y}{c}\right)^3} + b \hfill \\
    \Dfrac{y}{t} = z \hfill \\
    \Dfrac{z}{t} = - A\,x - \kappa\,y - \beta\,z - \kappa_{1}\,y^2
      - \kappa_{2}\,xy + B \hfill
  \end{gathered}
  \label{eq:thin-wall-limit_system}
\end{equation}
were $x = x(t) \equiv C(x,t) - C_{th}$ is the average concentration of
$Ca^{2+}$ in the arterial smooth muscle layer, $y = y(t) \equiv \eta(t)$
is the deviation of the radius of the vessel ($y > -c$),
$z=z(t)$ is the velocity of radius oscillation;
$\alpha$ is the rate of a natural ''pumping'' of the free calcium ions from
the intracellular space, $a$ is represents the blood flow level ($a \sim Q$),
$c$ is the non-perturbed arterial radius, $b$ is the rate of the calcium
inflow in a smooth muscle cell, $A$ is the coefficient of proportionality
for the calcium-feedback force, $\kappa$ is the linear elasticity coefficient,
$\kappa_{1}$ is the nonlinear elasticity coefficient, $\kappa_{2}$ is the small
calcium-induced elasticity coefficient, $\beta$ is the viscous (resistance)
coefficient of an arterial wall, $B$ represents the mean constant transmural
pressure.

Look for the stationary points of the system (\ref{eq:thin-wall-limit_system}).
One can obtain under the condition
\begin{equation}
  b - a = \alpha\frac{B}{A}
  \label{eq:constants_relation}
\end{equation}
there is a stationary point $\{x = B/A,\, y = 0,\, z = 0\}$. It corresponds to
the non-perturbed state of an artery. All the rest real stationary points of
the system have $y < -c$ and hence they are out of physical sense. The relation
(\ref{eq:constants_relation}) reflects the balance between the muscle forces
mediated by calcium concentration and the pressure forces in the blood.
The steady-state $Ca^{2+}$ concentration is equal to
$x = B/A \sim P_{0}/(h_{0} R_{0})$.

Study the stability of the dynamical system (\ref{eq:thin-wall-limit_system})
near the stationary point $\{B/A, 0, 0\}$ taking into account relation
(\ref{eq:constants_relation}). Consider the linearized system
\begin{equation}
  \Dfrac{\vec{X}}{t} = \mathbb{A} \vec{X} + \vec{F}\,,
\end{equation}
were
\begin{equation*}
  \begin{gathered}
  \mathbb{A} = \left(
  \begin{array}{ccc}
    -\alpha &            \frac{3 a}{c}           &       0  \\
          0 &                        0           &       1  \\
         -A &  -(\kappa + \kappa_{2}\frac{B}{A}) &  -\beta
  \end{array}
  \right)   \\
  \vec{X} = (x,\,y,\,z)^{T},\quad \vec{F} = (b-a,\, 0,\, B)^{T}
  \end{gathered}
\end{equation*}

The Routh-Hurwitz criterion provides the condition then all
eigenvalues of $\mathbb{A}$ have negative real parts. Here the
stability condition is as follows:
\begin{equation}
  \beta \left( \alpha^2 + \alpha\beta + \kappa + \kappa_{2}\frac{B}{A} \right)
      > \frac{3 a A}{c}
  \label{eq:stability_criterion}
\end{equation}

Taking into consideration the strictly positiveness of the $A, a, c, \alpha$
and non-negative values of the rest parameters one can conclude from
(\ref{eq:stability_criterion}) the condition for the wall viscosity
${\beta > 0}$. It shows the importance of the viscoelastic nature of
an arterial wall to maintain the stability of the stationary state.
In general case, there is the critical wall viscosity $\beta_{critical}$
below that oscillations demonstrate the lack of stability.

\begin{figure*}[!ht]
  \centering              
    \includegraphics[width=8cm]{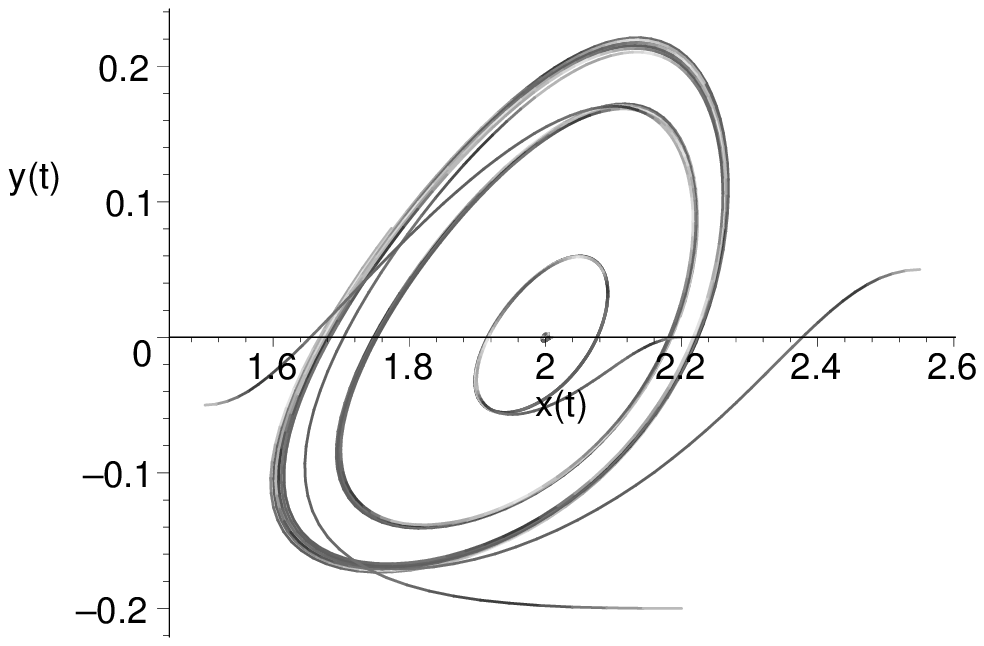}
    \includegraphics[width=8cm]{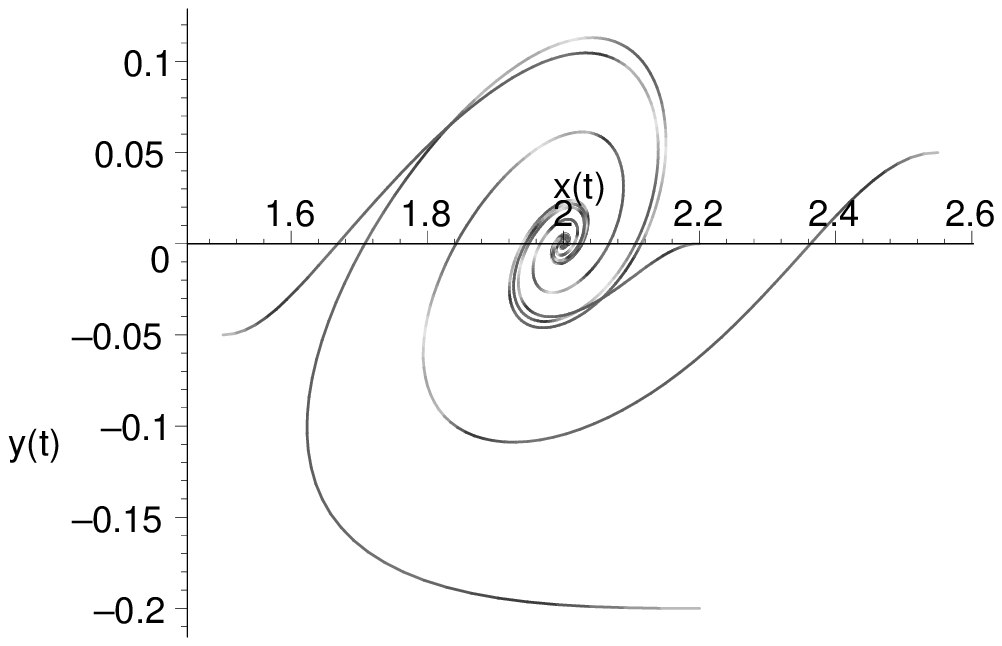}
    \caption{The two-dimensional projection of the phase trajectory
             of the system. For $\beta = \beta_{critical}$ it is the periodic
             oscillations (left) and for $\beta > \beta_{critical}$ it is the
             damping oscillations (right).}
  \label{fig:phase_plane}
\end{figure*}

The qualitative analysis on the phase plane confirms the preliminary estimates
(figure~\ref{fig:phase_plane}).

\section{The case of a passive vessel}\label{sec:Passive_tube}
One can see from the thin-wall approximation the more the discharge
the less the equilibrium calcium level. In the general model we have a
non-constant distribution of $Ca^{2+}$ concentration. If the stationary
$Ca^{2+}$ concentration for the whole arterial wall is below the threshold
level $C_{th}$ it becomes a fully relaxed. In this case the 'active'
viscoelastic tube is reduced to the 'passive' one. The law of the
arterial wall motion (\ref{eq:non-dim_wall_movement_eq}) in the
dimensionless form (primes are omitted) is as follows:
\begin{equation}
  \Dfrac[2]{\eta}{t} + \lambda\,\Dfrac{\eta}{t}
    + \varkappa_0\,\eta + \varkappa_1\,\eta^{2} = P_{0}
  \label{eq:non_dim_passive_wall_movement_eq}
\end{equation}

The nonlinear differential equation (\ref{eq:non_dim_passive_wall_movement_eq})
can be solved exactly via the simplest equation method
\cite{Kudryashov2005, Kudryashov1990}.
One can obtain
\begin{eqnarray}
  \label{eq:exact_passive-tube_solution}
  \eta(t) = \eta_{\infty} \tanh \left( \frac{\lambda\,t}{10} \right)
     \left( 2 - \tanh\left( \frac{\lambda\,t}{10} \right) \right)  \hfill  \\
  \eta_{\infty} = \sqrt{ \frac{P_0}{3\,\varkappa_{1}} },\quad
  \varkappa = \sqrt{ \frac{4\,\varkappa_{1}\,P_0}{3} },\quad
  \lambda = \sqrt[4]{ \frac{2500\,\varkappa_{1}\,P_0}{27} }  \nonumber \hfill
\end{eqnarray}

The kink-shaped solution demonstrates the switch from one steady state to
another under a constant force field.

Here the solution (\ref{eq:exact_passive-tube_solution}) satisfies
a non-perturbed state of artery with $\eta(0) = 0$. The pressure and
smooth-muscle force compensate each other. After a vanishing of the
muscle force (due to the sharp decreasing of the calcium level) arteria
expands to a new equilibrium state.

The new arterial radius depends on the transmural pressure and the elastic
properties of an arterial wall. It can be estimated by $\eta_{\infty}$.

\section{The numerical simulation for the problem of blood flow autoregulation}
\label{sec:Numerical_simulation}
Consider the general case of the two-layer kinetic-diffusion system in the
dimensionless form
(\ref{eq:non-dim_Ca_balance})~--~(\ref{eq:non-dim_boundary_conditions}) for
description of the blood flow regulation. In order to study the dynamics of the
solutions of the system near the steady state the numerical simulation is
performed. An implicit iterative finite-difference scheme
is implemented. As the initial values
the perturbed exact stationary solutions
(\ref{eq:steady-state_Ca}),~(\ref{eq:exact_steady-state_NO-distr}) are taken.

\begin{figure}[!ht]
  \centering              
    \includegraphics[width=8cm]{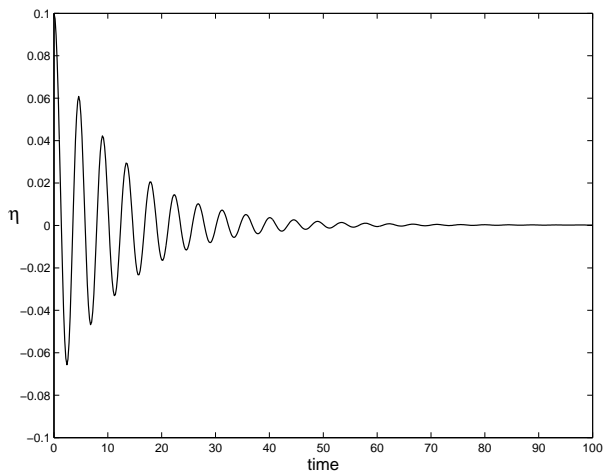}
    \caption{The dynamics of the system relaxation to the previous steady state
             after an initial stretching of the artery $\eta(t=0) = 0.1$.}
  \label{fig:eta_perturbed}
\end{figure}

The behavior of the solution for the initial stretching of the radius
confirms the asymptotic stability of the stationary state
(figure \ref{fig:eta_perturbed}).

As a test solution in case of passive dilation the exact solution
(\ref{eq:exact_passive-tube_solution}) is taken.

\begin{figure}[!ht]
  \centering              
    \includegraphics[width=8cm]{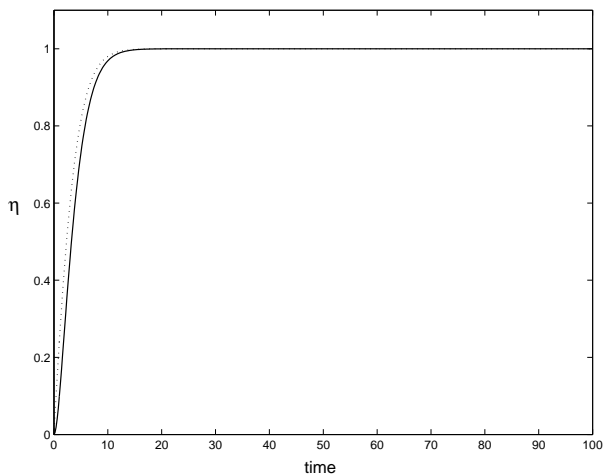}
    \caption{The passive expanding of the arteria due to a constant
             transmural pressure. The exact solution
             (\ref{eq:exact_passive-tube_solution}) (dotted line) and
             the numerical one (solid line).}
  \label{fig:test_case-passive_tube}
\end{figure}

The comparison gives a good agreement between the numerical solution and
the exact one (figure~\ref{fig:test_case-passive_tube}).

\begin{figure*}[!ht]
  \centering
    \includegraphics[width=8cm]{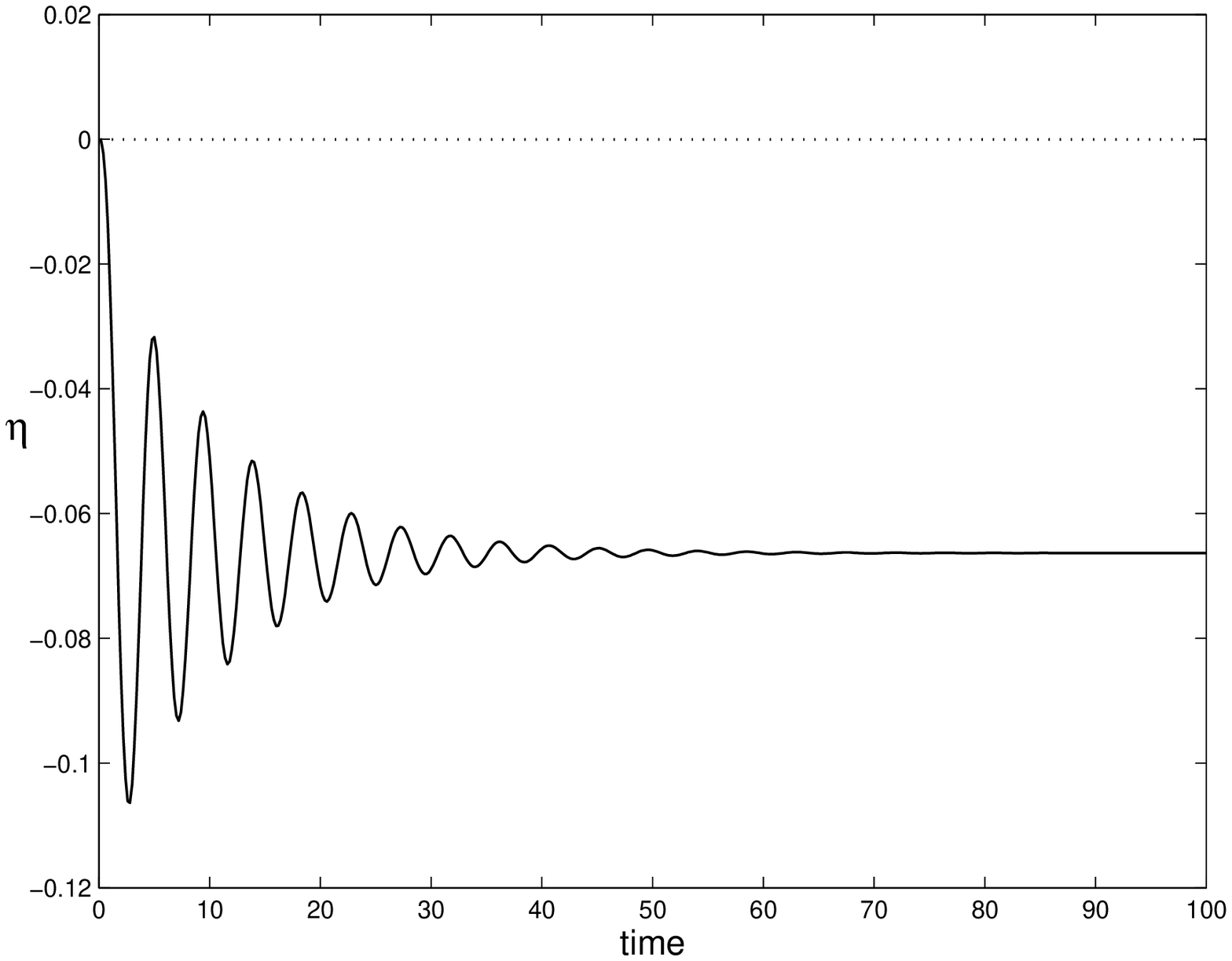}
    \includegraphics[width=8cm]{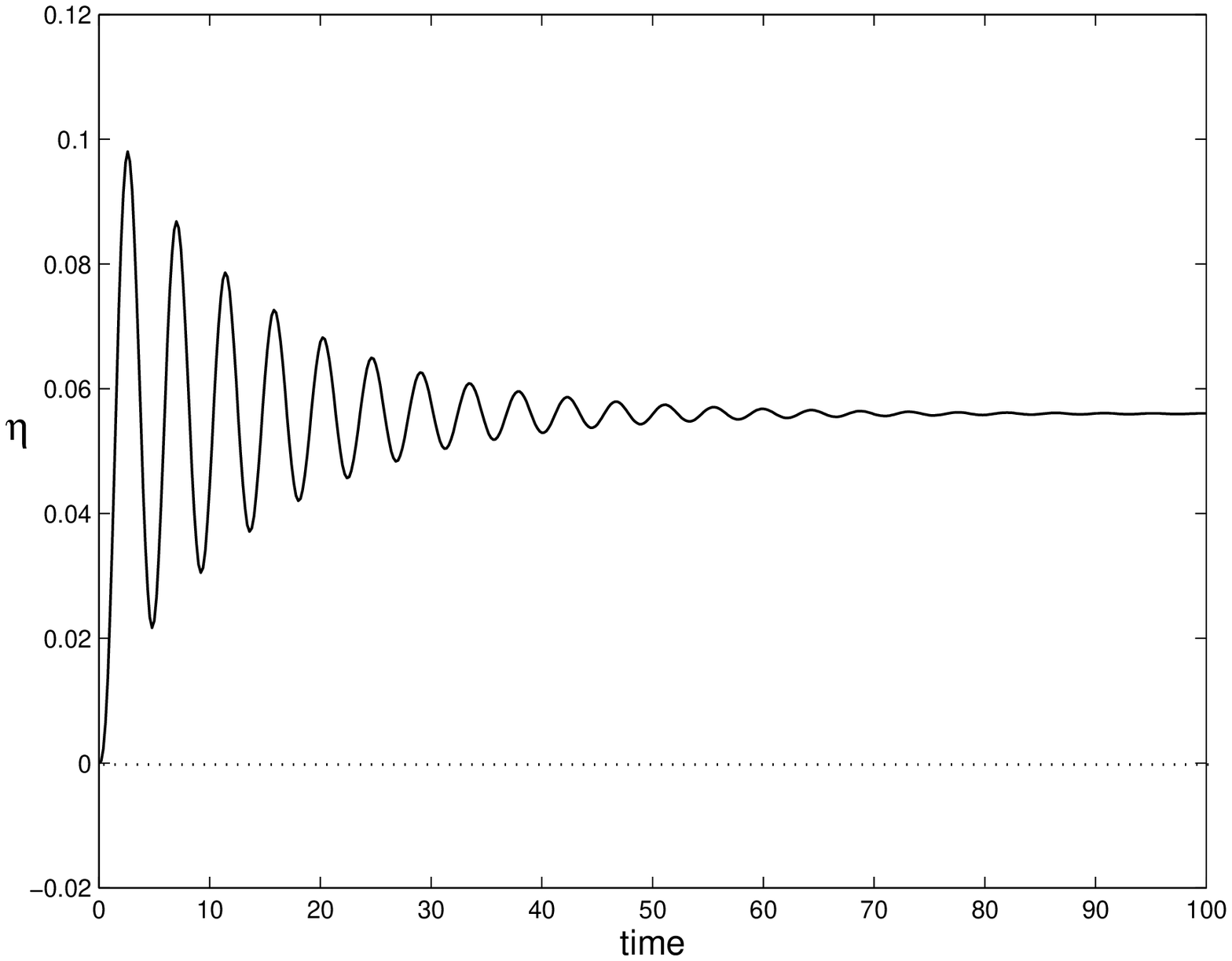}
    \caption{The transition of the system to the new equilibrium state after the
             decreasing for $25\%$ (left) and increasing for $25\%$ (right) of the
             mean blood flow.}
  \label{fig:flow_changing}
\end{figure*}

In response on the changing of the blood flow, that is managed by
the coefficient $k_{1} \sim Q$, the system comes after the damping
oscillations to a new steady state (figure \ref{fig:flow_changing}).
It is remarkable the different reaction of the system to the
increasing and decreasing of the discharge. The relaxation time in
case of flow decreasing is smaller than in case of flow increasing.
It may be explained by the drop of the critical viscosity level in
response on the decreasing flow according to
(\ref{eq:stability_criterion}). Also the arterial radius deviation
is bigger in case of decreasing of blood flow accordingly to the
inverse cubic dependence of the shear stress on the radius. In case
of increasing flow it is vice versa.

One can see the growth of the blood flow can potentially be a source of
instability especially for small arterial wall viscosity near the critical one.

\section{Conclusion}\label{sec:Conclusion}
The two-layer diffusion-kinetic model is proposed to describe the process of
a local blood flow regulation in an artery. The exact stationary distributions
of the key agents -- Nitric Oxide and Calcium ions are obtained.

The limit case of a thin wall artery under the estimation of the
wall thickness (\ref{eq:thickness_estimation}) is analytically
studied. The stability condition for the equilibrium state is given
by the formula (\ref{eq:stability_criterion}). It is shown the
necessity of the viscoelastic nature (non-zero viscosity) of the
arterial wall to provide the stability of the system. The minimal
critical viscosity value of a wall is obtained in the linearized
case.

In case of full relaxation of the smooth muscles the exact solution
in the kink form is found to describe the passive dilation of the
artery.

The numerical simulation demonstrates the transition of the system to the new
steady state with the new radius value in response of changing of the mean
blood discharge. This result is in agreement with the experimental
observation~\cite{Snow_etal2001}. It is confirmed the importance of the
endothelium derived relaxing factor -- Nitric Oxide for arterial haemodynamics.

The model can be applied to the study of the local autoregulation of the 
coronary, cerebral and kidney blood flow.

\begin{acknowledgments}
This work was supported by the International Science and Technology
Center under project B1213.
\end{acknowledgments}

\appendix
\section{The summary of notation}\label{sec:Appendix-notation}
Assuming the cylindrical coordinate system $( r, \theta, x \equiv z )$. \\
$R_{i}$ is the coordinate of intima layer boundary (equal to inner vessel
radius); \\
$R_{m}$ is the coordinate of media layer boundary; \\
$R_{a}$ is the coordinate of adventitia layer boundary; \\
$R=R(t)$ is the radius of the arterial wall ($R = R_{i}$
is the inner radius); \\
$\eta=\eta(t)=R(t)-R_0$ is the perturbation of the steady-state arterial radius $R_0$; \\
$C=C(r,t)$ is the concentration of $Ca^{2+}$-ions in the smooth muscle cell; \\
$C_{th}$ is threshold concentration of $Ca^{2+}$ to start the contraction in
the smooth muscle cell; \\
$n_{1}=n_{1}(r,t)$ is the concentration of $NO$-radical in the first (intima) layer; \\
$n_{2}=n_{2}(r,t)$ is the concentration of $NO$ in the second (media) layer; \\
$P_0=\bar{P} - P_{ext}=const$ is the cross-section averaged stationary transmural pressure; \\
$\bar{u}$ is the cross-sectional averaged axial fluid velocity of a steady-state flow; \\
$Q = A\,\bar{u}= const$ is the fluid discharge through a
cross-section of artery.

\section{Finding exact solution of the nonlinear
ODE (\ref{eq:non_dim_passive_wall_movement_eq})}
\label{sec:Appendix-SimpleEq-method}

To obtain an exact solution of the equation:
\begin{equation}
  \Dfrac[2]{y}{z} + \lambda\,\Dfrac{y}{z}
    + \varkappa_0\,y + \varkappa_1\,y^{2} = P_{0}
  \label{eq:nODE}
\end{equation}
we use the simplest equation method~\cite{Kudryashov2005} that
generalizes the existing approaches like the tanh-method, the method
of trial elliptic functions~\cite{Kudryashov1990}, etc.

Taking into account the second order pole of the general solution of
(\ref{eq:nODE}) look for solution in the form of the following expansion:
\begin{equation}
  y(z) = A_{0} + A_{1}\,G(z) + A_{2}\,G(z)^2
  \label{eq:expansion_for_y}
\end{equation}
where $G(z)$ is the solution with the first order pole of the equation
\begin{equation}
  \Dfrac{G(z)}{t} = k\,G(z) - k\,G(z)^2
  \label{eq:simplest_eq}
\end{equation}
and $A_0, A_1, A_2, k$ are the arbitrary constants to be determined.

Substituting the expansion (\ref{eq:expansion_for_y}) into equation
(\ref{eq:nODE}) we find
\begin{equation}
  \begin{gathered}
  A_0 =-\frac{30\lambda k + 25\varkappa - \lambda^2 + 25 k^2}{50\varkappa_1}\,,
    \quad
  A_1 = \frac{6 k (\lambda + 5 k)}{5 \varkappa_1}\,,  \\
  A_2 = -\frac{6 k^2}{\varkappa_1}\,,  \quad
  P_0 = -\frac{-36\lambda^4 + 625\varkappa^2}{2500\varkappa_1}\,,  \quad
  k = \pm\frac{\lambda}{5}
  \end{gathered}
\end{equation}

Taking the solution $G(z) = \frac{1}{2} + \frac{1}{2}
\tanh(\frac{1}{2}k(z-z_0))$ of the auxiliary equation
(\ref{eq:simplest_eq}) and choosing $k = \frac{\lambda}{5}$ we have
the exact solution of the ODE (\ref{eq:nODE}) in the form:
\begin{equation}
  \begin{gathered}
  y(z) = \frac{1}{50\varkappa_1} \Biglb( 3\lambda^2 - 25\varkappa
         + 6\lambda^2 \tanh\!\left[\tfrac{1}{10}\lambda(z-z_0)\right] - \\
         -3\lambda^2 \tanh^2\!\left[\tfrac{1}{10}\lambda(z-z_0)\right] \Bigrb)
  \label{eq:kink_solution}
  \end{gathered}
\end{equation}
where $z_0$ is an arbitrary constant.

Under the additional condition $y(0) = 0,\; z_0 = 0$ there are the following relations
between the parameters:
\begin{equation}
  \varkappa = \sqrt{ \frac{4\,\varkappa_{1}\,P_0}{3} },\quad
  \lambda = \sqrt[4]{ \frac{2500\,\varkappa_{1}\,P_0}{27} }
  \label{eq:additional_relations}
\end{equation}

After the simplification of (\ref{eq:kink_solution}) taking into account
(\ref{eq:additional_relations}) we obtain finally the kink-shape solution
of the equation (\ref{eq:nODE}):
\begin{equation}
   y(z) = y_{\infty} \tanh \left( \frac{\lambda\,z}{10} \right)
     \left( 2 - \tanh\left( \frac{\lambda\,z}{10} \right) \right)
\end{equation}
where $y_{\infty} = \sqrt{\frac{P_0}{3\,\varkappa_{1}}}$.


\end{document}